\documentclass[aps,pra,twocolumn,floatfix,10pt,nofootinbib]{revtex4}
\usepackage{amsmath,mathrsfs,graphicx}
\usepackage{multirow}
\begin{document}
\title{Measuring the Stokes parameters for light transmitted by a high-density rubidium vapour in large magnetic fields}
\author{Lee Weller, Toryn Dalton, Paul Siddons, Charles S Adams and Ifan G Hughes}
%\address{Department of Physics, Durham University, South Road, Durham, DH1~3LE, UK}
\affiliation{Department of Physics, Durham University, South Road, Durham, DH1~3LE, UK}
%\ead{lee.weller@durham.ac.uk}
\date{\today}

%%%%%%%%%%%%%%%%%%%%%%%%%%%%%%%%%%%%%%%%%%%%%%%%%%%%%%%%%%%%%%%%%%%%%%%%%%%%%%%%%%%%%%%%%%%%%%%%%%%%%%%%%%%%%%%%%%%%
\begin{abstract}
\noindent 
Here we report on measurements of the absolute absorption and dispersion of light in a dense rubidium vapour on the D$_{2}$ line in the weak-probe regime with an applied magnetic field.  A model for the electric susceptibility of the vapour is presented which includes both dipole-dipole interactions and the Zeeman effect.  The predicted susceptibility is comprehensively tested by comparison to experimental spectra for fields up to 800~G.  The dispersive properties of the medium are tested by comparison between experimental measurements and theoretical prediction of the Stokes parameters as a function of the atom-light detuning.
\end{abstract}
\maketitle
%%%%%%%%%%%%%%%%%%%%%%%%%%%%%%%%%%%%%%%%%%%%%%%%%%%%%%%%%%%%%%%%%%%%%%%%%%%%%%%%%%%%%%%%%%%%%%%%%%%%%%%%%%%%%%%%%%%%
\section{Introduction}
\label{Introduction}
%%%%%%%%%%%%%%%%%%%%%%%%%%%%%%%%%%%%%%%%%%%%%%%%%%%%%%%%%%%%%%%%%%%%%%%%%%%%%%%%%%%%%%%%%%%%%%%%%%%%%%%%%%%%%%%%%%%%
The study of light propagation through thermal atomic vapours subject to external magnetic fields is a flourishing area of research especially with a view to application in quantum information processing~\cite{hammerer2010quantum}.  Faraday~\cite{Faraday1846} discovered the magneto-optical phenomenon which bears his name where an axial magnetic field causes the rotation of the plane of polarization of incident linearly polarized light.  Much attention has been paid to resonant linear and nonlinear magneto-optical effects in atoms, and a comprehensive review of this field can be found in the article by Budker {\it et al}~\cite{budker2002resonant}.  The resonant atom-light interaction can be exploited to construct some of the most sensitive magnetometers~\cite{budker2007optical}.  The off-resonant Faraday effect is also beneficial in, for example, achieving a narrow-bandwidth optical filter with ultrahigh background rejection (the Faraday anomalous dispersion optical filter, FADOF) with atoms~\cite{dick1991ultrahigh} and crystals~\cite{lin2011proposed}; realizing a dichroic beam splitter for Raman light~\cite{abel2009faraday}; non-invasive atomic probing~\cite{siddons2009gigahertz}; and far off-resonance Faraday locking~\cite{marchant2011off}.

A theoretical model for the electric susceptibility of a vapour of alkali-metal atoms finds utility in, for example, analysing EIT spectra~\cite{gea1995electromagnetically,fleischhauer2005electromagnetically,mohapatra2007coherent}; understanding Doppler-broadened absorption spectroscopy~\cite{shin2010doppler}; designing a broadband optical delay line~\cite{vanner2008broadband,camacho2007wide}; controlling ultrabroadband slow light~\cite{zhang2011controllable}; enhancing the frequency up-conversion of light~\cite{vernier2010enhanced}; and achieving quantitative spectroscopy for a primary standard~\cite{truong2011quantitative}.  In our group we have developed a model for the absolute susceptibility that allows us to make quantitative predictions for the absorptive and dispersive properties in the vicinity of the D lines\footnote{The transitions $n^{2}$S$_{1/2}$~$\rightarrow$~$n^{2}$P$_{1/2}$ and $n^{2}$S$_{1/2}$~$\rightarrow$~$n^{2}$P$_{3/2}$, where $n$ is the principal quantum number of the valence electron, are referred to as the D$_{1}$ and D$_{2}$ transitions, respectively, for alkali-metal atoms.}.  The model has successfully accounted for absolute Doppler-broadened absorption in the low-density regime in rubidium (Rb)~\cite{siddons2008absolute} and cesium (Cs)~\cite{kemp2011analytical}; absolute absorption including dipole-dipole interactions in the binary-collision regime in Rb~\cite{weller2011absolute}; and dispersion using the Faraday effect~\cite{siddons2009off}.       

The motivation of this study is to include the effect of an axial magnetic field up to 800~G in the model for susceptibility and to measure the modification to the optical absorptive and dispersive properties of the atoms.  Note that the field is sufficiently large that the angular momentum $F$ is partially uncoupled for the ground terms, and $F'$ and $m_F'$ are not good quantum numbers for the excited states.  The absorptive properties are comprehensively tested through measuring experimental Doppler-broadened absorption spectra for both low-density and binary-collision regimes.  In addition the dispersive properties are investigated by measuring the Stokes parameters of the transmitted light.  
 
The structure of the remainder of the paper is as follows.  In section~\ref{Experimental Method}, we describe the experimental methods.  In section~\ref{High-Density Rubidium vapour in a Large Magnetic Field} we modify the electric susceptibility to incorporate an axial magnetic field, and show absolute absorption spectra.  In section~\ref{Stokes parameters and the Poincare representation} we compare experimental and theoretical Stokes parameters for off-resonant light transmitted through a high-density Rb vapour.  In addition we represent the evolution of the polarization state of light as a function of detuning on the Poincar\'{e} sphere.  Finally in section~\ref{Conclusions} we summarize our findings. 
%%%%%%%%%%%%%%%%%%%%%%%%%%%%%%%%%%%%%%%%%%%%%%%%%%%%%%%%%%%%%%%%%%%%%%%%%%%%%%%%%%%%%%%%%%%%%%%%%%%%%%%%%%%%%%%%%%%%
\section{Experimental Method}
\label{Experimental Method}
\begin{figure}[t]
\centering
\includegraphics*[width=0.45\textwidth]{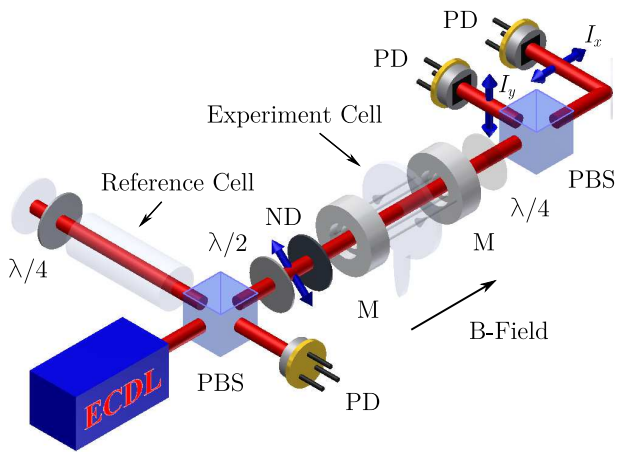}
\caption{Schematic of the experimental apparatus.  A beam passes through a polarization beam splitter (PBS), providing linearly polarized light.  A small fraction of the beam is used to perform hyperfine/saturated absorption spectroscopy in a room-temperature reference cell.  A half-wave plate ($\lambda$/2) is set at $-\pi/8$ rads to the linearly polarized light before it is separated into its horizontal and vertical components and collected on each photodiode (PD).  The probe beam is attenuated with a neutral-density filter (ND) before passing through a heated experiment cell and two magnets (M).  The magnetic field is applied axially along the direction of the beam and an optional quarter-wave plate ($\lambda$/4) is available for measuring the Stokes parameters.}
\label{MainSetup}
\end{figure}

%%%%%%%%%%%%%%%%%%%%%%%%%%%%%%%%%%%%%%%%%%%%%%%%%%%%%%%%%%%%%%%%%%%%%%%%%%%%%%%%%%%%%%%%%%%%%%%%%%%%%%%%%%%%%%%%%%%%
Figure~\ref{MainSetup} shows a schematic of the experimental apparatus used to observe the modification to the absorptive and dispersive properties of the atoms in the presence of a magnetic field.  An external cavity diode laser system was used for these measurements with a wavelength of 780~nm and scanned across the Rb D$_{2}$ transition.  The laser output passes through a polarization beam splitter providing linearly polarized light with a 1/e$^2$ radius of 0.68~$\pm$~0.03~mm.  A Fabry-Perot etalon (not shown) was used to linearize the frequency scan with a small fraction of the beam passing through a natural-abundant room temperature reference cell performing hyperfine/saturated absorption spectroscopy~\cite{macadam1992narrow,smith2004role} to calibrate the scan.  A half-wave plate ($\lambda$/2) is set at $-\pi/8$ rads to the linearly polarized light such that in the absence of any optical rotation the differencing signal is zero, as the light intensities of the horizontal, $I_{x}$, and vertical, $I_{y}$, channels of light incident on the individual detectors are equal~\cite{HuardPolarization1997}.  A neutral-density filter then attenuates the probe beam before it traverses a 2~mm heated experiment cell containing Rb in its natural abundance (72$\%$ $^{85}$Rb, 28$\%$ $^{87}$Rb).  For powers much less than 100~nW, the atoms in the cell traversing the probe beam do not undergo hyperfine pumping into the other ground term hyperfine level and good agreement between theory and experiment is obtained~\cite{siddons2008absolute,sherlock2009weak,kemp2011analytical}.    

The cell was placed in an oven with the same design as in~\cite{weller2011absolute}.  An aluminum holder was used to control the magnitude of the magnetic field by varying the separation between two countersunk Neodymium magnets which have central holes for the laser beam to pass through.  The magnetic field profile is measured with a Hall probe.  Over the length of the cell the field is uniform at the 1\% level.  After traversing the experiment cell and magnet holder a polarization beam splitter separates the light into horizontal and vertical components before the light impinges on calibrated photodiodes.  To observe all four Stokes parameters a quarter-wave plate ($\lambda$/4) must also be inserted.   

Transmission spectra were measured in the absence and presence of a large magnetic field, in both the low-density and binary-collision regime.  The Stokes parameters were measured with the laser red detuned from the D$_{2}$ $^{87}$Rb $F$ = 2 $\rightarrow$ $F'$ = 1,\,2,\,3 transitions in the binary-collision regime for a modest magnetic field.
%%%%%%%%%%%%%%%%%%%%%%%%%%%%%%%%%%%%%%%%%%%%%%%%%%%%%%%%%%%%%%%%%%%%%%%%%%%%%%%%%%%%%%%%%%%%%%%%%%%%%%%%%%%%%%%%%%%%
\section{High-Density Rubidium vapour in a Large Magnetic Field}
\label{High-Density Rubidium vapour in a Large Magnetic Field}
%%%%%%%%%%%%%%%%%%%%%%%%%%%%%%%%%%%%%%%%%%%%%%%%%%%%%%%%%%%%%%%%%%%%%%%%%%%%%%%%%%%%%%%%%%%%%%%%%%%%%%%%%%%%%%%%%%%%
\subsection{Theory}
%%%%%%%%%%%%%%%%%%%%%%%%%%%%%%%%%%%%%%%%%%%%%%%%%%%%%%%%%%%%%%%%%%%%%%%%%%%%%%%%%%%%%%%%%%%%%%%%%%%%%%%%%%%%%%%%%%%%
The (complex) electric susceptibility, $\chi$, encapsulates both the dispersive and absorptive properties of a medium.  In a Doppler-broadened atomic vapour, $\chi$ for the transition $j$ is given by~\cite{siddons2008absolute}
\begin{align}
\chi_{j}\left(\Delta\right)=c_{j}^{2}\frac{d^{2}\mathcal{N}}{\hbar\epsilon_{0}}s_{j}\left(\Delta\right),
\label{susceptibility}
\end{align}
where $\Delta$ is the detuning from resonance (the difference in angular frequency between the laser and the atomic resonance); $c_{j}^2$ is the transition strength; $d$ is the reduced dipole matrix element; $\hbar$ is the reduced Planck constant; $\epsilon_0$ is the permittivity of free-space and $\mathcal{N}$ is the (temperature dependent) atomic number density.  $s_{j}\left(\Delta\right)$ is the lineshape of the resonance; for an atomic vapour this is a convolution of a Lorentzian lineshape (accounting for natural and self broadening~\cite{weller2011absolute}) and a Gaussian distribution to include the Doppler shift arising from the component of motion along the propagation direction of the laser.  The total susceptibility for the D$_{2}$ line is obtained by summing over the electric-dipole allowed transitions among all the hyperfine sub-levels.

For each hyperfine level, $F$, there are $2F + 1$ magnetic sublevels which are degenerate in energy in the absence of an external magnetic field.  The atomic Hamiltonian can be written as 
\begin{align}
\hat{H} = \hat{H}_0 + \hat{H}_{\rm fs} + \hat{H}_{\rm hfs} + \hat{H}_{\rm Z}~,
\label{eq:hamilton}
\end{align}
where $\hat{H}_0$ is describes the coarse atomic structure; $\hat{H}_{\rm fs}$ and $\hat{H}_{\rm hfs}$ are the fine and hyperfine interactions; and $\hat{H}_{\rm Z}$ represents the atomic interaction with an external magnetic field.  For Rb the transition frequencies and frequency intervals associated with $\hat{H}_0 + \hat{H}_{\rm fs} + \hat{H}_{\rm hfs}$ are known precisely; the numerical values can be found in, for example, Table~1 of reference~\cite{siddons2008absolute}.
 
The magnetic interaction Hamiltonian for an external field $\boldsymbol{\cal B}$, which defines the  $z-$axis, is of the form
\begin{align}
\hat{H}_{\rm Z} = - (\boldsymbol{\mu}_I + \boldsymbol{\mu}_J) \cdot \boldsymbol{\cal B}~, 
\end{align}
where $\boldsymbol{\mu}_I$ and $\boldsymbol{\mu}_J$ are the magnetic moments of the nucleus and electron, respectively.  As the nuclear magneton is three orders of magnitude smaller than the Bohr magneton we shall ignore the contribution of the nuclear magnetic moment~\cite{ramsey1956molecular}.  In the absence of field the $\vert F, m_F\rangle$ basis is appropriate.  The hyperfine splitting for the $^2S_{1/2}$ terms of $^{87}$Rb and $^{85}$Rb are 6.8 and 3.0~GHz~\cite{siddons2008absolute}, respectively, therefore the hyperfine sublevels undergo a linear Zeeman shift for fields of up to a few hundred Gauss.  For higher fields the Zeeman shift is nonlinear.  For the excited $^2P_{3/2}$ terms of Rb the hyperfine intervals vary between 30 and 270~MHz~\cite{siddons2008absolute} and for fields exceeding $\sim$10~G the Zeeman shift is nonlinear. 

In order to calculate the contributions to the susceptibility of equation~(\ref{susceptibility}) at any given field value a numerical approach is adopted.  A matrix representation of $\hat{H}_{\rm fs}+\hat{H}_{\rm hfs} + \hat{H}_{\rm Z}$ is calculated in the completely uncoupled $\vert m_I,\,m_{\ell},\,m_s\rangle$ basis.  Here $m_I,\,m_{\ell},$ and $m_s$ are the projections of the nuclear spin, the electronic orbital angular momentum and the electron's spin, respectively.  Both the magnetic dipole and electric quadrupole contributions to the hyperfine Hamiltonian~\cite{ramsey1956molecular} are incorporated.  Numerical diagonalisation of the resulting matrix yields the frequency detunings and the optical transition strengths are calculated subject to the selection rules $\Delta m_I =0$, $\Delta m_s =0$, and $\Delta m_{\ell}=\pm1$ for $\sigma^{\pm}$ transitions, respectively.  When calculating the D$_{2}$ spectrum we use the magnetic dipole and electric quadrupole coefficients for the $^{2}P_{3/2}$ term -- this is an excellent approximation for the fields of interest to us in this work where the external field has magnitude $\leq 1$~kG, and the Zeeman shift is four orders of magnitude smaller than the fine-structure splitting. 
%%%%%%%%%%%%%%%%%%%%%%%%%%%%%%%%%%%%%%%%%%%%%%%%%%%%%%%%%%%%%%%%%%%%%%%%%%%%%%%%%%%%%%%%%%%%%%%%%%%%%%%%%%%%%%%%%%%%%%%%%%%%%%%%%%%%%%%%%%%%%%%%%%%%%%%%%%%%%%%%%%%
\subsection{Experimental Results}
\begin{figure}[t]
\centering
\includegraphics*[width=0.45\textwidth]{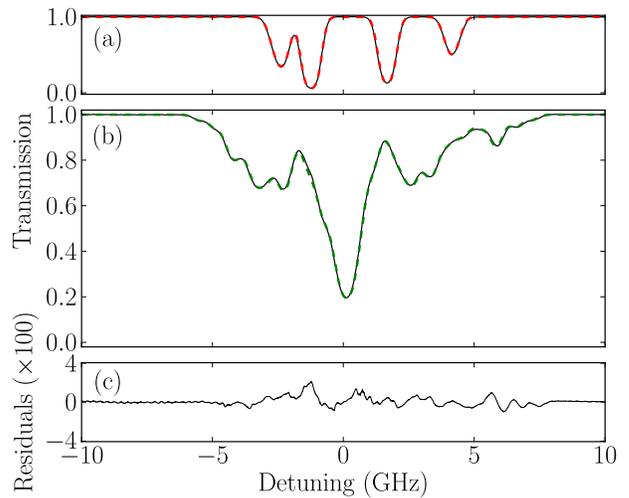}
\caption{Transmission plots for the comparison between experiment and theory for the Rb D$_{2}$ line, through a vapour cell of length 2~mm as a function of linear detuning, $\Delta/2\pi$.  Zero detuning is taken to be the centre-of-mass frequency of the transition.  Plots (a) and (b) show the transmission in the absence and presence of a magnetic field, respectively.  Solid (black) and dashed (red and green) lines show measured and expected transmission, for a Lorentzian width of $\Gamma_{0}/2\pi$~=~6.065~MHz~\cite{volz1996precision}, with a temperature of $(76.7~\pm~0.1)$~$^\circ$C in a magnetic field of $(766~\pm~5)$~G.  Plot (c) shows the residuals (the difference in transmission between experiment and theory) for (b).  There is excellent agreement between the data and model, with an rms deviation of 0.4$\%$.}
\label{HighFieldLowTemp}
\end{figure}

%%%%%%%%%%%%%%%%%%%%%%%%%%%%%%%%%%%%%%%%%%%%%%%%%%%%%%%%%%%%%%%%%%%%%%%%%%%%%%%%%%%%%%%%%%%%%%%%%%%%%%%%%%%%%%%%%%%%%%%%%%%%%%%%%%%%%%%%%%%%%%%%%%%%%%%%%%%%%%%%%%
Figure~\ref{HighFieldLowTemp} shows a plot of the transmission of the Rb D$_{2}$ line versus linear detuning, $\Delta/2\pi$.  The zero of the detuning axis is taken to be the centre-of-mass frequency of the transition in the absence of hyperfine splitting, taking into account the natural abundance of each isotope.  The solid (black) lines show the transmission measured by one of the photodiodes before the PBS in (a) the absence, and (b) the presence of a magnetic field, at the same temperature.  The dashed (red and green) lines are the corresponding theoretical transmission spectra using the susceptibility of~\cite{siddons2008absolute}.  The two theory curves are generated with a Lorentzian width of $\Gamma_0$ (the natural width) and a Doppler width which is allowed to vary.  A least-squares fit~\cite{MATU} allows us to extract the temperature.  Five spectra are recorded for each temperature to allow the statistical variation to be quantified; for the parameters which were used to generate figure~\ref{HighFieldLowTemp} the fits yield a temperature of $(76.7~\pm~0.1)$~$^\circ$C, which is consistent with a thermocouple measurement.  In figure~\ref{HighFieldLowTemp}~(b) the presence of the field means that the spectrum is very rich; there are many spectral features, all of which are accounted for in the theoretical model.  Figure~\ref{HighFieldLowTemp}~(c) shows the residuals which demonstrate that there is excellent agreement between the data and model with an rms deviation of 0.4$\%$.  Five spectra are recorded for each value of magnetic field, and for the parameters of figure~\ref{HighFieldLowTemp}~(b) we extract $(766~\pm~5)$~G which is consistent with measurements with a Hall probe.  For this magnitude of field the angular momentum $F$ is partially uncoupled for the ground terms, and $F'$ and $m_F'$ are not good quantum numbers for the excited states -- this is the origin of the numerous spectral features in figure~\ref{HighFieldLowTemp}~(b).
 
\begin{figure}[t]
\centering
\includegraphics*[width=0.45\textwidth]{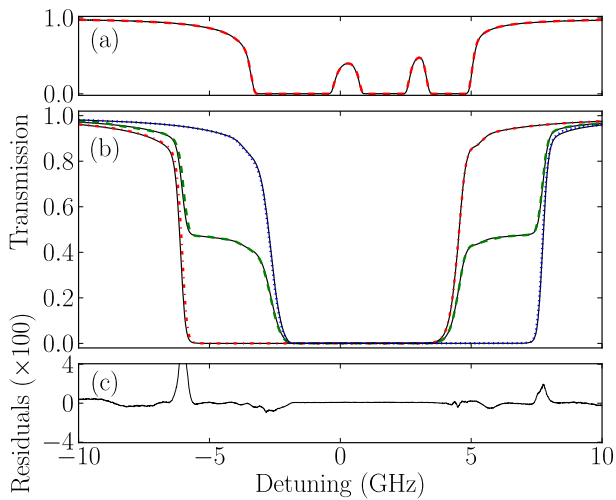}
\caption{Transmission plots for the comparison between experiment and theory for the Rb D$_{2}$ line, through a vapour cell of length 2~mm as a function of linear detuning, $\Delta/2\pi$.  Plots (a) and (b) show the transmission including resonant dipole-dipole interactions, in the absence and presence of an applied magnetic field, respectively.  The three spectra are the total transmission (dashed green) and the transmission of the right (dot-dashed red) and left (dotted blue) circular polarization components.  The fits yield a Lorentzian width of $\Gamma/2\pi$~=~$(23.3~\pm~0.4)$~MHz, with a temperature of $(159.8~\pm~0.2)$~$^\circ$C in a magnetic field of $(774~\pm~4)$~G.  Plot (c) shows the residuals of the total transmission demonstrating that there is excellent agreement between the data and model (with the exception of a small number of glitches where the linearization of the laser scan was not adequate), with an rms deviation of 0.7$\%$.}
\label{HighFieldHighTemp}
\end{figure}

Figure~\ref{HighFieldHighTemp} shows the transmission spectrum of the Rb D$_{2}$ line at elevated temperature where dipole-dipole interactions become important~\cite{weller2011absolute} and the Lorentzian width has an additional density-dependent term as well as the natural contribution.  The solid (black) lines show the measured transmission in (a) the absence, and (b) presence of a magnetic field.  Three transmission spectra are recorded in the presence of field: the dashed (green) curve with a single photodiode placed before the PBS, and the dot-dashed (red) and dotted (blue) recored on the two photodiodes after a quarter waveplate is added to show the transmission of light driving $\sigma^{-}$ and $\sigma^{+}$ transitions, respectively.  The least-squares fits to the four theory curves yield a Lorentzian width of $\Gamma/2\pi$~=~$(23.3~\pm~0.4)$~MHz, a temperature of $(159.8~\pm~0.2)$~$^\circ$C in a magnetic field of $(774~\pm~4)$~G for figure~\ref{HighFieldHighTemp}~(b).  There is excellent agreement between the data and model except for some minor glitches in the residuals where the transmission varies most rapidly, an rms deviation of 0.7$\%$ was measured.  Note that the absorption spectra of figure~\ref{HighFieldHighTemp}~(b) have simpler profiles than the lower-temperature spectrum of \ref{HighFieldLowTemp}~(b).  In particular on either side of the resonance there are spectral windows approximately 4~GHz wide where one circular polarization component is nearly fully transmitted and the orthogonal component nearly fully absorbed.  A thorough investigation of the performance of a filter exploiting the circular dichroism in the wings of the absorption spectrum will be the subject of a future publication. 

%%%%%%%%%%%%%%%%%%%%%%%%%%%%%%%%%%%%%%%%%%%%%%%%%%%%%%%%%%%%%%%%%%%%%%%%%%%%%%%%%%%%%%%%%%%%%%%%%%%%%%%%%%%%%%%%%%%%
\section{Stokes parameters and the Poincar\'{E} representation}
\label{Stokes parameters and the Poincare representation}
%%%%%%%%%%%%%%%%%%%%%%%%%%%%%%%%%%%%%%%%%%%%%%%%%%%%%%%%%%%%%%%%%%%%%%%%%%%%%%%%%%%%%%%%%%%%%%%%%%%%%%%%%%%%%%%%%%%%
We can represent the polarization state of transmitted light via the Stokes parameters~\cite{stokes1852polarization,schaefer2007measuring}.  Combinations of the observable intensities of various polarization components allow us to measure the four Stokes parameters, which are defined as:
\begin{align}
S_{0}& = I_{-} + I_{+} = I_{0} \frac{1}{2} \left(e^{-\alpha^{-} L} + e^{-\alpha^{+} L}\right)~,\\
S_{1}& = I_{x} - I_{y} = I_{0} \cos\left(2\varphi\right)e^{-\frac{1}{2}\left(\alpha^{-}+\alpha^{+}\right)L}~,\\
S_{2}& = I_{\nearrow} - I_{\searrow} = I_{0} \sin\left(2\varphi\right)e^{-\frac{1}{2}\left(\alpha^{-}+\alpha^{+}\right)L}~,\\
S_{3}& = I_{-} - I_{+} = I_{0} \frac{1}{2} \left(e^{-\alpha^{-} L} - e^{-\alpha^{+} L}\right)~.  
\end{align}
The parameter $S_{0}$ describes the total intensity of the transmitted light field.  $S_{1}$ describes the intensity difference between horizontal and vertical linearly polarized light.  $S_{2}$ describes the intensity difference between linearly polarized light at an angle $+\pi/4$ rad and $-\pi/4$ rad to the $x$-axis.  $S_{3}$ describes the intensity difference between right- and left-circularly polarized light.  $I_{0}$ is the initial intensity and $\varphi$ is the rotation angle with respect to the $x$-axis.  Note that the parameter $S_{0}$  can be the sum of the intensity components in any of the three described bases.  After propagating through a medium of length $L$ the left- and right- circularly polarized components of the light field experience a phase shift $\phi^{\pm} = k_0 n^{\pm} L$.  The rotation angle is therefore $\varphi = \frac{1}{2}\left(\phi^{+} - \phi^{-}\right) + \theta_{0} = \theta + \theta_{0}$, where for balanced polarimetry the initial rotation angle $\theta_{0}$ is set to $-\pi/4$ rad, such that the $S_{1}$ parameter tends to zero when $\theta$ tends to zero.

\begin{figure}[b]
\centering
\includegraphics*[width=0.45\textwidth]{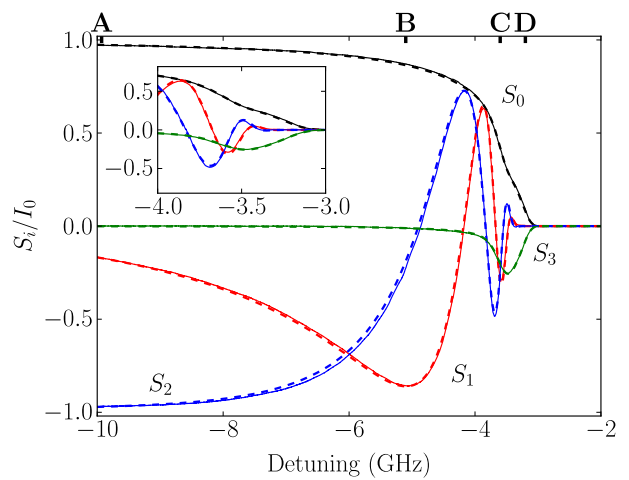}
\caption{Comparison between experimental and theoretical Stokes parameters for the Rb D$_{2}$ line, through a vapour cell of length 2~mm as a function of linear detuning, $\Delta/2\pi$, red detuned from the D$_{2} ^{87}$Rb $F$ = 2 $\rightarrow F'$ = 1,\,2,\,3 transitions.  Solid and dashed lines show measured and expected $S_{0}$ (black), $S_{1}$ (red), $S_{2}$ (blue) and $S_{3}$ (green) signals, for a Lorentzian width of $\Gamma/2\pi$~=~$(23.3~\pm~0.4)$~MHz, at a temperature of $(159.8~\pm~0.2)$~$^\circ$C in a magnetic field of $(82~\pm~2)$~G.  \textbf{A},\,\textbf{B},\,\textbf{C} and \textbf{D} are four different frequencies showing how the polarization state of light evolves as a function of detuning.  The inset gives detail of the evolution close to resonance.}
\label{StokesParameters}
\end{figure}

Figure~\ref{StokesParameters} shows the comparison between experimental and theoretical Stokes parameters for a Doppler-broadened medium of natural Rb atoms with a Lorentzian width of $\Gamma/2\pi$~=~$(23.3~\pm~0.4)$~MHz, at a temperature of $(159.8~\pm~0.2)$~$^\circ$C in a magnetic field of $(82~\pm~2)$~G, red detuned from the D$_{2}$ $^{87}$Rb $F$ = 2 $\rightarrow$ $F'$ = 1,\,2,\,3 transitions.  Far detuned from resonance (\textbf{A}) the absorptive and dispersive properties of Rb are negligible.  As the light is tuned closer to resonance (\textbf{B}) dispersive properties dominate the light-matter interaction, leading to birefringence and a varying $S_{1}$ and $S_{2}$ signal. As $S_{0}$ and $S_{3}$ depend only on the absorption, and not the refractive index of the medium, they are both very similar to their extreme far from resonance values. Closer to resonance (\textbf{C})  absorptive properties become apparent, circular dichroism expresses itself resulting in a change in the $S_{3}$ signal.  Even closer to resonance the right-circular component is completely absorbed (\textbf{D}) because the $\sigma^-$ transition is at a higher red-detuning than the $\sigma^+$ transition and the $S_{1}$ and $S_{2}$ signals stop varying.  At this detuning the transmitted light has left-circular polarization.  For detunings closer still to resonance the medium is optically thick to both polarizations and all four Stokes parameters are zero.  There is excellent agreement between the experimental and theoretical Stokes parameters, which demonstrates that our theoretical model accounts successfully for both the absorptive and dispersive components of the atom-light interaction in a large magnetic field.  Any discrepancy is likely due to the different detectors used in the measurement of $I_{x}$ and $I_{y}$ and the calibration of the frequency axis.    

\begin{figure}[t]
\centering
\includegraphics*[width=0.45\textwidth]{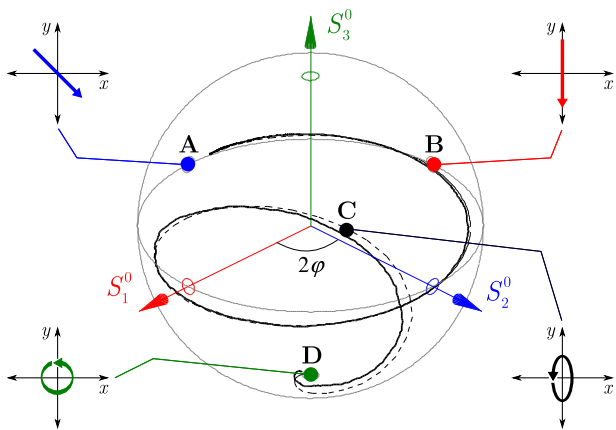}
\caption{Representation of the polarization state of light on the Poincar\'{e} sphere.  The normalized Poincar\'{e} vector ($S^0_{1}$,\,$S^0_{2}$,\,$S^0_{3}$) is measured and calculated for a Doppler-broadened medium of natural Rb atoms with a Lorentzian width of $\Gamma/2\pi$~=~$(23.3~\pm~0.4)$~MHz, at a temperature of $(159.8~\pm~0.2)$~$^\circ$C in a magnetic field of $(82~\pm~2)$~G, red detuned from the D$_{2}$ $^{87}$Rb $F$ = 2 $\rightarrow$ $F'$ = 1,\,2,\,3 transitions.  The poles represent left- and right-circular polarization, the equator linear polarization and intermediate points elliptical polarization.  \textbf{A},\,\textbf{B},\,\textbf{C} and \textbf{D} are four different frequencies showing how the polarization state of light evolves as a function of detuning.}
\label{PoincareSphere}
\end{figure}   

The Poincar\'{e} sphere is a graphical way to represent the polarization state of light~\cite{HuardPolarization1997}.  For polarized light we can present a normalized Poincar\'{e} vector ($S^0_{1}$,\,$S^0_{2}$,\,$S^0_{3}$) where $S^0_{i}$ = $S_{i}/S_{0}$ for $i = 1,2,3$. The south and north poles denote left- and right-circularly polarized light, respectively; linearly polarized light is mapped to the equator; on every other point on the sphere the light is elliptically polarized.  Figure~\ref{PoincareSphere} shows a normalized Poincar\'{e} sphere measured and calculated for a Doppler-broadened medium of natural Rb atoms with a Lorentzian width of $\Gamma/2\pi$~=~$(23.3~\pm~0.4)$~MHz, at a temperature of $(159.8~\pm~0.2)$~$^\circ$C in a magnetic field of $(82~\pm~2)$~G, red detuned from the D$_{2}$ $^{87}$Rb $F$ = 2 $\rightarrow$ $F'$ = 1,\,2,\,3 transitions.  (\textbf{A}) shows the initial linearly polarized state at an angle $-\pi/4$ rad to the $x$-axis.  As the light is tuned closer to resonance (\textbf{B}) the polarization state of the light initially remains linear whilst experiencing a rotation around the equator of the sphere.  Closer to resonance (\textbf{C}) the polarization state moves off the equator and becomes highly elliptical, before becoming purely left-circularly polarized at the southern hemisphere (\textbf{D}).  The magnetic field chosen for the data sets of figures~\ref{StokesParameters} and \ref{PoincareSphere} is an order of magnitude smaller than that used to generate the data for figures~\ref{HighFieldLowTemp} and \ref{HighFieldHighTemp}.  The motivation for this is the clarity of figure~\ref{PoincareSphere}.  For a 774~G field there would be many rotations around the vertical axis of the Poincar\'{e} sphere.  The excellent agreement between the experimental and theoretical trajectories on the Poincar\'{e} sphere, confirm the accuracy of the predictions of the absorptive and dispersive properties of our model for the electric susceptibility of the dense atomic vapour including an external field.
%%%%%%%%%%%%%%%%%%%%%%%%%%%%%%%%%%%%%%%%%%%%%%%%%%%%%%%%%%%%%%%%%%%%%%%%%%%%%%%%%%%%%%%%%%%%%%%%%%%%%%%%%%%%%%%%%%%%
\section{Conclusions}
\label{Conclusions}
%%%%%%%%%%%%%%%%%%%%%%%%%%%%%%%%%%%%%%%%%%%%%%%%%%%%%%%%%%%%%%%%%%%%%%%%%%%%%%%%%%%%%%%%%%%%%%%%%%%%%%%%%%%%%%%%%%%%
In summary we have discussed the physics underlying the transmission of light through a dense atomic vapour, accounting for self-broadening and the application of a large axial magnetic field.  We showed excellent agreement between experimental and theoretical absorption spectra of rubidium vapour on the D$_{2}$ line, and of the spectral dependence of the Stokes parameters red detuned from the D$_{2}$ $^{87}$Rb $F$ = 2 $\rightarrow$ $F'$ = 1,\,2,\,3 transitions.  These results demonstrate that our theoretical model of the electric susceptibility of the atomic vapour accounts successfully for both the absorptive and dispersive components of the atom-light interaction in a large magnetic field.  The Poincar\'{e} sphere was shown to be a useful representation of the spectral dependence of the polarization of a probe beam transmitted through the medium.  In future we will characterize the performance of a filter exploiting the extreme circular dichroism in the wings of the absorption spectrum.
%%%%%%%%%%%%%%%%%%%%%%%%%%%%%%%%%%%%%%%%%%%%%%%%%%%%%%%%%%%%%%%%%%%%%%%%%%%%%%%%%%%%%%%%%%%%%%%%%%%%%%%%%%%%%%%%%%%%
\section*{Acknowledgements}
%\ack
%%%%%%%%%%%%%%%%%%%%%%%%%%%%%%%%%%%%%%%%%%%%%%%%%%%%%%%%%%%%%%%%%%%%%%%%%%%%%%%%%%%%%%%%%%%%%%%%%%%%%%%%%%%%%%%%%%%%
This work is supported by EPSRC.  Toryn Dalton was funded by the Ogden Trust.      
%%%%%%%%%%%%%%%%%%%%%%%%%%%%%%%%%%%%%%%%%%%%%%%%%%%%%%%%%%%%%%%%%%%%%%%%%%%%%%%%%%%%%%%%%%%%%%%%%%%%%%%%%%%%%%%%%%%%
%\nocite{*}
%\bibliographystyle{iopart-num}
%\bibliographystyle{unsrt}
\bibliography{myreferences}
%%%%%%%%%%%%%%%%%%%%%%%%%%%%%%%%%%%%%%%%%%%%%%%%%%%%%%%%%%%%%%%%%%%%%%%%%%%%%%%%%%%%%%%%%%%%%%%%%%%%%%%%%%%%%%%%%%%%
\end{document}